\documentclass[a4paper,reqno]{amsart}
\usepackage{geometry}
\usepackage{fullpage}
\usepackage[latin1]{inputenc}
\usepackage[T1]{fontenc}
\usepackage{amsfonts}
\usepackage{amssymb}
\usepackage{amsmath}
\usepackage{amsthm}
\usepackage{graphicx}
\usepackage[dvipsnames]{xcolor}
\usepackage{afterpage}
\usepackage[colorlinks=true, linkcolor=magenta, citecolor=cyan, urlcolor=green]{hyperref} 
\usepackage{bbm}
\usepackage{latexsym}

\newcommand{\rme}{{\rm e}}

\newcommand{\bb}{\mbox{\boldmath $b$}}

\newcommand{\bsigma}{\mbox{\boldmath $\sigma$}}
\newcommand{\btau}{\mbox{\boldmath $\tau$}}

\numberwithin{equation}{section}

\definecolor{light}{gray}{.9}

\def\be{\begin{equation}}
\def\ee{\end{equation}}
\def\bea{\begin{eqnarray}}
\def\eea{\end{eqnarray}}

\def\s{\sigma}

\def\e{\varepsilon}
\def\epsilon{\e}

\newcommand{\nocontentsline}[3]{}
\newcommand{\tocless}[2]{\bgroup\let\addcontentsline=\nocontentsline#1{#2}\egroup}

\DeclareMathSymbol{\leqslant}{\mathalpha}{AMSa}{"36} 
\DeclareMathSymbol{\geqslant}{\mathalpha}{AMSa}{"3E} 
\DeclareMathSymbol{\eset}{\mathalpha}{AMSb}{"3F}     
\renewcommand{\leq}{\;\leqslant\;}                   
\renewcommand{\geq}{\;\geqslant\;}                   

\title{Retrieving infinite numbers of patterns in a
spin-glass model of immune networks}
\date{\today}

\author{Elena Agliari}

\address{Elena Agliari: Dipartimento di Matematica, Sapienza Universit\`a di Roma, Italy.}
\email{elena.agliari@mat.uniroma1.it}

\author{Alessia Annibale}

\address{Alessia Annibale: Department of Mathematics, King's College University of London, UK.}
\email{alessia.annibale@kcl.ac.uk}

\author{Adriano Barra}

\address{Adriano Barra: Dipartimento di Matematica e Fisica Ennio De Giorgi, Universit\`a del Salento, Italy.}
\email{adriano.barra@unisalento.it}

\author{ A.C.C. Coolen}

\address{ A.C.C. Coolen:  Institute for Mathematical and Molecular Biomedicine, King's College University of London (UK).}
\email{ton.coolen@kcl.ac.uk}

\author{Daniele Tantari}

\address{Daniele Tantari: Scuola Normale Superiore, Centro Ennio de Giorgi, Italy.}
\email{daniele.tantari@sns.it}

\begin{document}

\begin{abstract}
 The similarity between neural and (adaptive) immune networks has been known for decades, but so far we did not  understand the mechanism that allows the immune system, unlike associative neural networks,  to recall and execute a large number of memorized defense strategies {\em in parallel}. The explanation turns out to lie in the network topology. Neurons interact typically with a large number of other neurons, whereas interactions among lymphocytes in immune networks are very specific, and described by graphs with finite connectivity.
In this paper we use replica techniques to solve  a statistical mechanical immune network model with `coordinator branches' (T-cells) and `effector branches' (B-cells),  and show how  the finite connectivity enables the coordinators to manage an extensive number of effectors  simultaneously, even above the percolation threshold (where clonal cross-talk is not negligible).
\newline
A consequence of its underlying topological sparsity is that the adaptive immune system exhibits only weak ergodicity breaking, so that also spontaneous switch-like effects as bi-stabilities are present: the latter may play a significant role in the maintenance of immune homeostasis.
\end{abstract}


\maketitle

Beyond the so-far-classical approaches by Cohen, DeBoer, May, Nowak and Perelson  (see e.g. \cite{zerox,zerozero,zerouno,zerodue,zerotre}) that paved the {\em main route} for mathematical modelling in Immunology, and after a pioneering early paper by Parisi \cite{parisi} followed by about two decades of dormancy, there is now increasing interest in statistical mechanical approaches to modeling the immune system \cite{JTB1,JTB2,bialek,kosmir1,kosmir2,prlnoi,deem,silvia1,silvia2,ton6}. This interest is stimulated in part by the potential of new quantitative methods for the study of systems with complex network topologies \cite{barabasi,prlTon,ton3,ton4,smallworld}. In this paper we show how statistical mechanics can resolve a central problem in theoretical immunology:  understanding the parallel processing ability of the subclass of lymphocytes that are dedicated to the coordination of the adaptive immune response, i.e. helper and regulator T-cells.

T- and B-lymphocytes are divided into clones. Cells of the same B clone  detect and attack the same antigens, and are selected for activation when their allocated antigens invade the host.
Conditional on authorization by T-helpers (via eliciting cytokines), the selected B-cells undergo clonal expansion: they multiply, and start releasing high quantities of soluble antibodies to inhibit the enemy. After the antigen has been deleted, B-cells are no longer triggered, thus -instructed by T-regulators (via suppressive cytokines)- stop producing antibodies and undergo apoptosis. In this way the clones reduce their sizes, and order is restored. We stress that two signals are required for B-cell clones to expand. The first arises from antigen binding; the second is a `consensus' signal, a cytokine secreted by T-helpers. This AND-gate like mechanism \cite{SR1,SR2} prevents abnormal reactions, such as autoimmunity \cite{janaway,JTB1}.
The core of the immune adaptive response thus consists of an effector branch (the B-clones \footnote{The effector branch includes also e.g. killer T-cells \cite{janaway},  which will not be considered here for simplicity. See e.g. \cite{JTB1}.}) and a coordination branch (the helper and regulator T-clones), which interact through cytokines that convey either eliciting or suppressive signals. This can be modeled as a collection of interacting variables on a bipartite network, endowed with specific `spin-glass couplings' \cite{JTB1,JTB2} (see Fig.s $1a,\ 1b$.)

The immune system is able to learn (e.g. how to fight new antigens), memorize (e.g. previously seen antigens) and  `think'  (e.g. select the best strategy to deal with pathogens), all of which it has in common with neural networks. However, the architectures of neural and immune networks are very different.
Neurons tend to have a huge number of connections with others \cite{Ton} (for instance cortical modules in mammals are known to share hierarchical organization of densely connected clusters \cite{hierarchical1,hierarchical2}, far above the giant component appearance), thus overpercolated  network models (mathematically convenient) are more tolerable in the neural scenario. In contrast, the interactions among lymphocytes (via chemical messengers, i.e. cytokines) are very specific and short range: the underlying topology displays finite connectivity. This difference  plays a crucial operational role \cite{peter-prl,gerarchico-prl,long}. Neural network models perform high-resolution serial processing, which is achieved by many spins (neurons)  interacting extensively.  We will show that the immune system's  striking ability to cope with many antigens {\em simultaneously}, instead, can be understood as a direct consequence of  having many spins (lymphocytes) that interact in an intelligent {\em sparse} manner.
\begin{figure}[t]
\vspace*{1mm}
 \begin{center}
\hspace*{-2mm}
 (a)~
\includegraphics[width=33mm]{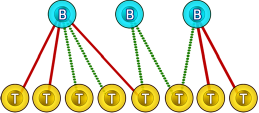} \ \ \ \
(b)~
\includegraphics[width=33mm]{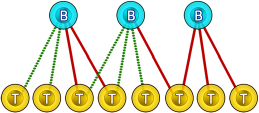}
\hspace*{2mm}
\end{center}
\vspace*{-4mm}
\begin{center}
\hspace*{-2mm}
(c)~
\includegraphics[width=33mm]{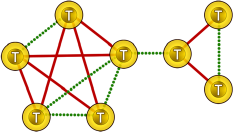} \ \ \ \
(d)~
\includegraphics[width=33mm]{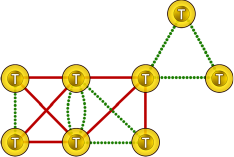}
\hspace*{2mm}
\end{center}
\vspace*{-2mm}
\caption{\label{fig:casi} Examples of connected components in the bipartite interaction graph $\mathcal{B}$ of the model (\ref{eq:fullH}) with interacting B and T-cells (upper panels),  and the corresponding connected components in the equivalent graph $\mathcal{G}$ of the effective  system (\ref{eq:reducedH}) with T-cells only (lower panels). Dashed green links mark positive interactions; solid red links mark negative ones. 
}
\end{figure}
\begin{figure}[t]
\vspace*{0.5mm}
 \begin{center}{
\includegraphics[width=.235\textwidth]{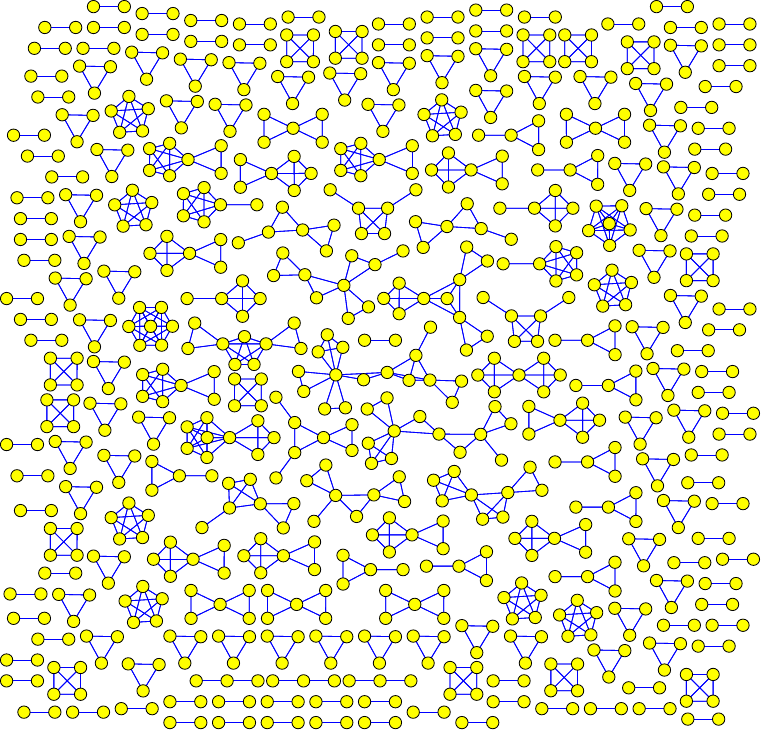}~~
\includegraphics[width=.235\textwidth]{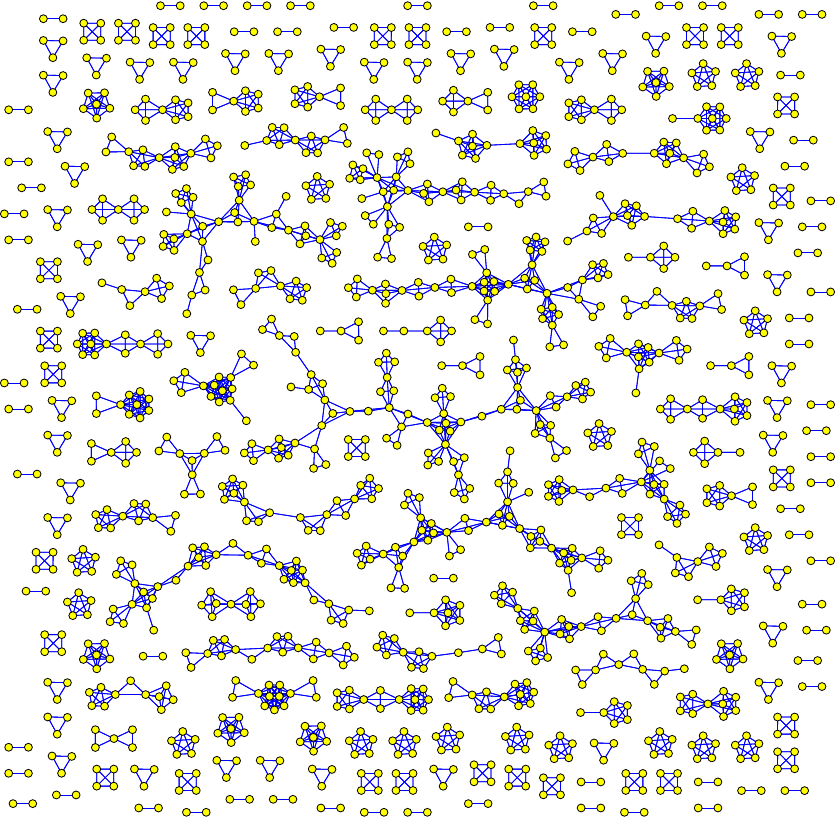}}
\end{center}
\vspace*{-2mm}
\caption{\label{fig:grafi2}\small Examples of graphs $\mathcal{G}$ obtained for different values of $c$, with $N_T=5000$ and $\alpha=0.1$. Recalling the critical threshold $\alpha c^2 =1$, here we compare graphs with $\alpha c^2 <1$ (below percolation, left panel) and $\alpha c^2=1$ (percolation threshold, right panel). Isolated nodes are not shown; their number is $4229$ and $3664$, respectively. As expected, loops between cliques start to occur at the percolation threshold.
}
\end{figure}

Let us consider an immune repertoire of $N_B$ different B-clones, labeled by $\mu \in \{1,...,N_B\}$. The size of  clone $\mu$ is $b_{\mu} \in \mathcal{R}$. In the absence of interactions with antigens and T-cells (i.e. {\em at rest}), we take clonal sizes to be Gaussian distributed; this is supported both by experiments and theoretical arguments \cite{JTB1}. Without loss of generality we may take zero means  and unit widths, i.e. $P(b_{\mu}) \sim \mathcal{N}(0,1)$. A value $b_{\mu}\gg 0$ then indicates that clone $\mu$ has expanded (relative to the typical clonal size), while $b_{\mu}\ll 0$ implies  inhibition.  As in standard reaction kinetics (where chemical potentials scale linearly with the fields, i.e. logarithmically with the concentrations, when framed in statistical mechanical terms \cite{thompson}),  the relation between the relative concentration of B cells and their clonal sizes is logarithmical (apart a constant factor that sets the proper scale, i.e. at rest the average clone size is of $O(10^3)$ \cite{janaway}), see \cite{noi2016} for  details. Similarly, we consider $N_T$ T-clones, labeled by $i \in \{1,...,N_T\}$.
The state of  T-clone $i$ is denoted by $\s_i$. For simplicity, T-clones are assumed to have just two possible states: secreting cytokines ($\s_i=+1$) or quiescent ($\s_i=-1$), see \cite{JTB1} for details.
The cytokine $\xi_i^{\mu}$ secreted by helper $i$ and detected by clone $\mu$ is described by a discrete variable, carrying either an excitatory ($\xi_i^{\mu} = +1$) or inhibitory ($\xi_i^{\mu} = -1$) instruction; the value, $\xi_i^\mu=0$ is used to indicate lack of signalling among clones $i$ and $\mu$. The pattern of cytokines, which describes the interactions between T and B clones, represents a  bipartite graph, denoted as $\mathcal{B}$. Its $N_T N_B$ entries $\{\xi_i^{\mu}\}$ are quenched\footnote{Cytokines are split into several families (e.g. interferons, interleukins) and here they are assumed to be quenched because they do not evolve over time \cite{janaway}; however, a more refined model should take into account a range of values broader than $\pm 1$ in order to capture their different strength.}, and taken to be  independently distributed according to
\be\label{chito}
P(\xi_i^\mu)= \frac{c}{2N_T}(\delta_{\xi_i^\mu,+1}+\delta_{\xi_i^\mu,-1})+(1-\frac{c}{N_T})\delta_{\xi_i^\mu,0}
\ee
with $c>0$.
As stated, we focus on the biologically relevant regime \cite{janaway}: finite connectivity, i.e. $c=\mathcal{O}(N_T^0)$, and high storage, i.e. $N_B = \alpha N_T$ with $\alpha>0$ fixed, while $N_B,\ N_T\to\infty$.  Here the number of B and T-clones are comparable and the interactions between cells do not scale with the system size, mirroring {\em chemical specificity};  further, as the amount of different clones is  of order $O(10^9)$, we assume that a theory developed in the thermodynamic limit (as the one we are presenting here) is somehow reasonable.

$P(\xi_i^{\mu})$ implicitly accounts for bond dilution in the graph  $\mathcal{B}$.
In particular, when the link probability $c/N_T$ exceeds the percolation threshold $1/\sqrt{N_T N_B}$, i.e. for $\alpha c^2 >1$, the graph $\mathcal{B}$ will have a giant component (see Fig. $2$).

To highlight the computational capabilities of such a system, as in the route paved in neural networks \cite{Ton,AGS}, in these first steps we restrict ourselves to an {\em equilibrium analysis}. Here the probability of a configuration $(\bb,\bsigma)$ is captured by the relative Gibbs weight $P(\bb,\bsigma) \propto \exp(-\beta \hat{\mathcal{H}}(\bb,\bsigma|\xi))$: we introduce an effective Hamiltonian $\hat{\mathcal{H}}(\bb,\bsigma|\xi)$ -that has no  mean in terms of the {\em energy of the system} as in the classical framework of statistical mechanics- at an inverse noise level $\beta \equiv 1/ T$ (where $T$, that in Physics plays the role of the temperature, is the proxy for the -standard/white- noise strength). In these regards, the usage of the Gibbs measure has to be understood under the Maximum Entropy Principle perspective \cite{jaynes,bialek1} (again, as standard in neural networks \cite{bialek2}, and as already started to be applied in theoretical immunology \cite{bialek}).
\newline
The {\em effective  Hamiltonian}  for the combined T and B-cell system \cite{prlnoi,AACTold}, interacting on the graph $\mathcal{B}$, reads as
\be
\hat{\mathcal{H}}(\bb,\bsigma|\xi) = -\frac{1}{\sqrt{c}}\sum_{i=1}^{N_T}\sum_{\mu=1}^{N_B}\xi_i^{\mu}\s_i b_{\mu} +\frac{1}{2\sqrt{\beta}}\sum_{\mu=1}^{N_B}b_\mu^2.
\label{eq:fullH}
\ee
In the language of Disordered Systems, this is a hyper-diluted bipartite spin-glass, while in the jargon of Machine Learning this is a Boltzmann machine with a Gaussian regularizer. Crucially, in the partition function $Z$, en route to the free energy and the system's thermodynamics, we can integrate out the $b_\mu$ \cite{prlnoi,AACTold}, viz.
\be\small
Z = \sum_{\bsigma}\int\!d\bb~e^{-\sqrt{\beta} \hat{\mathcal{H}}(\bb,\bsigma|\xi)}=\sum_{\bsigma}
e^{-\beta H(\bsigma|\xi)},
\ee
where $H(\bsigma|\xi)$ now includes T-T interactions only:
\be
H(\bsigma|\xi)= -\frac{1}{2c}\sum_{ij=1}^{N_T}\sum_{\mu=1}^{N_B}\xi_i^\mu\xi_j^\mu\sigma_i\sigma_j
= -\frac{1}{2c}\sum_{\mu=1}^{N_B}M_\mu^2(\bsigma).
\label{eq:reducedH}
\ee
Here $M_\mu(\bsigma) \equiv \sum_i\xi_i^\mu \sigma_i$ is the non-normalized overlap between the T-cell state $\bsigma$ and the vector $(\xi_1^\mu,\ldots,\xi_N^\mu)$.
The B-T system on the bipartite graph $\mathcal{B}$ has thereby been mapped to an equivalent effective T-T system on a monopartite weighted graph $\mathcal{G}$, in which the coupling between  node
pairs $(i, j)$ has the Hebbian form \cite{AGS,Ton} $J_{ij} = \sum_{\mu=1}^{N_B} \xi_i^{\mu} \xi_j^{\mu}$ (see Fig.s $\ 1c,\ 1d$). It follows that T-clones can retrieve stored cytokine signalling patterns. To understand  the immunological meaning of pattern retrieval, we focus on the B-clone $\mu$ and consider the case where each T-clone $i$ is `aligned' with the related cytokine $\xi_i^{\mu}$ (if nonzero). Those $i$ that inhibit clone $\mu$ (i.e. secrete $\xi_{i}^\mu=-1$)  will be quiescent ($\s_{i}=-1$), and those $i$ that excite $\mu$ (i.e. secrete $\xi_{i}^\mu=+1$)  will be active ($\s_{i}=+1$). This state gives the maximum of $M_\mu(\bsigma)$, i.e. of the overall signal received by B-clone $\mu$, see eq. (\ref{eq:fullH}): the random environment becomes a`staggered magnetic field' that  forces the expansion of clone $\mu$,  so the arrangement of T-cells leading to the retrieval of pattern $\mu$ corresponds to maximal clone-specific excitatory signalling upon  B-clone $\mu$.
If $\xi_i^\mu\in\{-1,1\}$ for all  $(i, \mu)$, so the bipartite network  is fully connected, retrieval will operate as in the Hopfield model \cite{AGS}; the system could expand only one B-clone at a time and this would be a disaster for immuno-surveillance. If the immune system is to manage an extensive number of expanded B-clones simultaneously, it will require extreme dilution.

Let us now take a topological perspective. We note that in the \emph{under-percolated} regime the graph $\mathcal{B}$ is a forest, where the typical components are (combinations of) stars centered on a $B$-node (because experimentally $N_T > N_B$ \cite{janaway}); see Fig.~\ref{fig:casi}a. Such trees are mapped into complete graphs
or combinations of complete graphs in $\mathcal{G}$ (Fig.~\ref{fig:casi}c). Therefore, when $\alpha c^2< 1$ the typical components in $\mathcal{G}$ are of finite size (see Fig.~$2$) and may  form cliques whose occurrence frequency decays exponentially with their size.
In this regime, two T-nodes $i, j $ have at most one common neighboring B-node $\mu$, so the  spins  $\sigma_i$ and $\sigma_j$ can propagate non-conflicting signals to $\mu$. We thus expect this regime to be compatible with parallel retrieval.
Parallel retrieval can be jeopardized by the presence of loops in $\mathcal{B}$, which create alternative feed-back routes between spins; see Fig.~\ref{fig:casi}b.
The probability that a loop occurs in $\mathcal{B}$ scales as $( \alpha c^2)^2$ \cite{long},
so loops should appear near the percolation threshold. In the graph $\mathcal{G}$, such a loop implies that two cliques can share not only nodes but also links, and that two T-nodes can have a coupling $|J_{ij}| \geq 2$ (see Fig.~\ref{fig:casi}d and  Fig.~$2$). As a result, the simultaneous retrieval of all patterns within the same component is no longer ensured.

\begin{figure}[t]
\unitlength=0.5mm
\hspace*{-3mm}
\begin{picture}(180,98)
\put(2,-5){\includegraphics[width=180\unitlength]{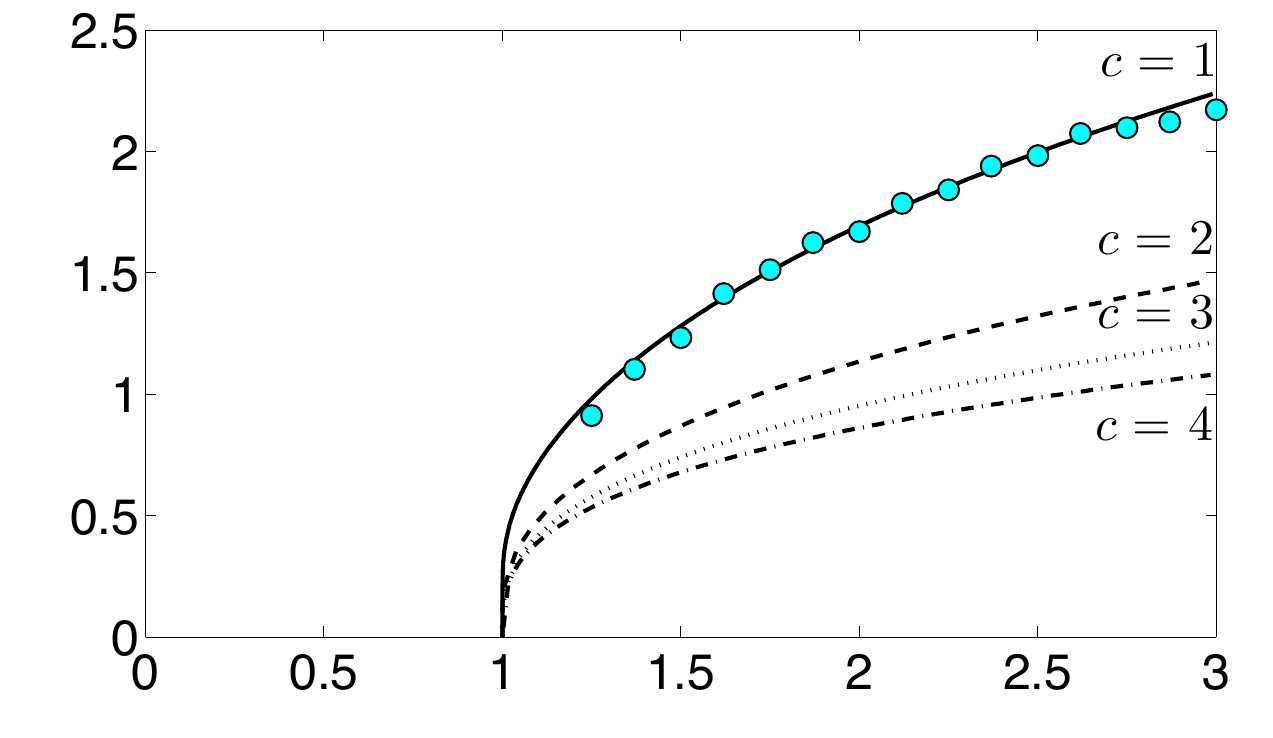}}
\put(80,-8){$\alpha c^2$}
\put(3,50){$T$}
\put(125,22){\rm clonal}
\put(125,15){\rm cross-talk}
\put(32,83){\rm parallel processing of}
\put(32,76){\rm extensively many clones}
\end{picture}
\vspace*{1mm}
\caption{Transition lines (\ref{eqn:crit}) for $c=1,2,3,4$, in the $(\alpha,T)$ plane. In the parallel processing phase
 the effective T-T network can successfully control an extensive number of B-clones simultaneously. In the clonal cross-talk phase (at low temperatures above the precolation point)  the connectivity causes interference between clone-specific strategies. Circles: transition calculated via numerical solution of (\ref{eq:single_eqn}) for $c=1$. 
 }
\label{fig:Tc}
\end{figure}
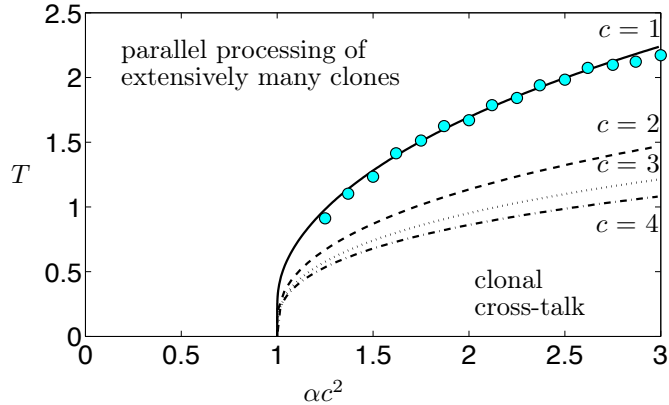

\begin{figure}[t]
\hspace*{-2.2mm}
\includegraphics[width=.48\textwidth]{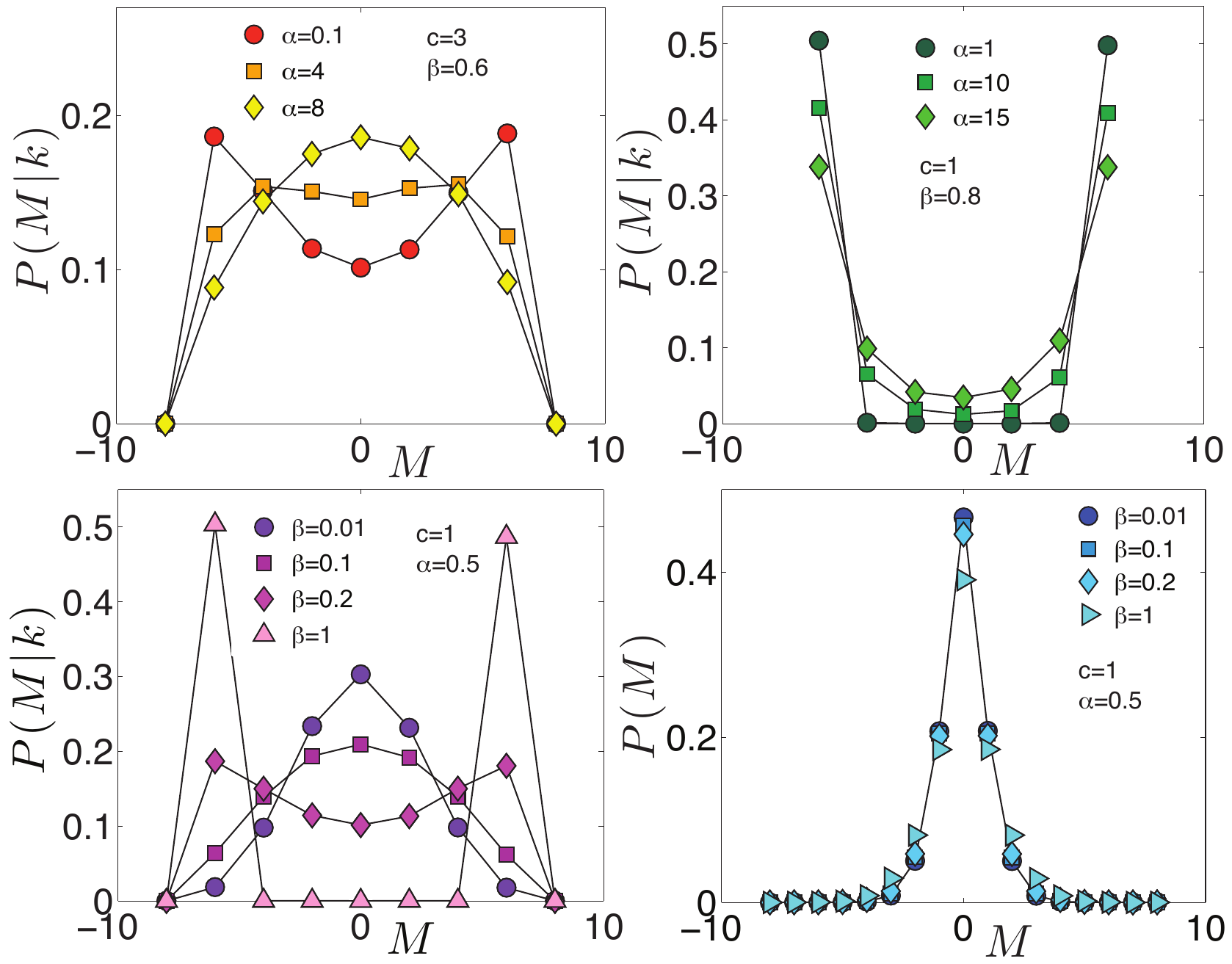}
\caption{Top panels: effect on $P(M|k)$ of moving into the cross-talk regime by increasing $\alpha$ (left: $c=3$ and $T=5/3$; right: $c=1$ and $T=5/4$). Bottom panels: shapes of $P(M|k)$ and $P(M)$ for $\alpha c^2=1/2$, in  the parallel processing regime. All values are calculated from the solution of (\ref{eq:single_eqn},
\ref{eqn:pm}), for $k=6$. 
} \label{fig:under}
\end{figure}

%

Hyper-dilution in $\mathcal{B}$ is apparently crucial for extensive multiple clonal expansions. It ensures that  patterns to be retrieved in $\mathcal{G}$ have many blank entries and that, unlike neural networks,  `pure states'  are no  longer low energy configurations. Retrieving a pattern  $(\xi_1^\mu,\ldots,\xi_N^\mu)$ does not involve all spins $\sigma_i$, and those corresponding to null entries  can be used to recall other patterns. This is energetically favorable since the energy (\ref{eq:reducedH}) is quadratic in the magnetizations $M_\mu(\bsigma)$. However, to quantify retrieval within this new scenario we need alternative (and more refined) order parameters beyond standard Mattis magnetizations. The distribution $P(M)= N_B^{-1}\sum_{\mu=1}^{N_B}\delta_{M,M_\mu(\bsigma)}$ of Mattis magnetizations  would work perfectly to the case, but it contains entangled information, from the thermal  magnetization fluctuations within a single pattern, and from fluctuations over different patterns.  Upon denoting with $P_c(k)$ the prior that a pattern has $k$ non-zero entries, we can  disentangle the different contributions by focusing on $P(M|k)$, the conditional magnetization distribution for patterns with $k$ nonzero entries, defined via  $P(M)=\sum_{k=1}^{N_T} P_c(k) P(M|k)$.
We can easily calculate $P_c(k) $, because it depends only on the structure of  $\mathcal{B}$. Since we have $N_T$ independent entries, each nonzero  with probability $c/N_T$, in the thermodynamic limit the variable $k$ is  Poissonian distributed:
\be
P(M)=\rme^{-c}\sum_{k=0}^{\infty} \frac{c^k}{k!}  P(M|k).
\label{eqn:pm}
\ee
With this observable we can in fact solve the present model analytically, and calculate the free energy per spin  using the finite connectivity replica method,  within the replica-symmetric approximation (RS). Full details of this (somewhat lengthy) calculation have been published elsewhere  \cite{AACTold,long}. The result leads to  an explicit expression for $P(M|k)$ in terms of an effective field distribution $W(h)$, which is to be solved in
a self-consistent way (see eq.s (\ref{eqn:pm}) and (\ref{eq:single_eqn})),
\begin{eqnarray}  
P(M|k) &=& \sum_{r\geq 0}
\frac{e^{-\alpha c k}(\alpha c)^r}{r!}
 \int   d \mathbf{h}   \cdot \\ \nonumber 
&& \prod_{s\leq r} W(h_s) \sum_{l_1\ldots l_r=1}^k \left\{ \frac{N_M(\tau|h)}{D_M(\tau|h)}
\right\},\label{eqn:pm} \\  
W(h) &=&
\sum_{k\geq 0}\frac{e^{-c}c^k }{k!} \sum_{r\geq 0}\frac{e^{-\alpha c k}(\alpha c)^r}{r!}
\int d \mathbf{h} \cdot \\ \nonumber 
&& \prod_{s\leq r}\!W(h_s) \sum_{l_1\ldots l_r=1}^k \langle \delta [h\!-\!\frac{1}{2\beta}\ln\frac{N_W(\tau|h)}
{D_W(\tau|h)}]
\rangle_{\tau}
\label{eq:single_eqn}
\end{eqnarray}
with the short-hand $\langle f(\tau)\rangle_{\tau}=\frac{1}{2}\sum_{\tau=\pm 1}f(\tau)$ and where
\begin{eqnarray} \nonumber
N_M(\tau|h)&=&  \big\langle
\delta_{M,\sum_{l\leq k}\tau_l}
 e^{\frac{\beta}{2c} (\sum_{l\leq k}\tau_l)^{2}+\beta \sum_{s\leq r} h_s \tau_{l_s}}
\big\rangle_{{{\btau}}} \\  \nonumber
D_M(\tau|h)&=&  \big\langle
\rme^{\frac{\beta}{2c} (\sum_{l\leq k}\tau_l)^{2}+\beta \sum_{s\leq r} h_s \tau_{l_s}}
\big\rangle_{{\btau}} \\  \nonumber
N_W(\tau|h)&=&
\langle
\rme^{\frac{\beta}{2c}(\sum_{l\leq k}\tau_{l})^2+\frac{\beta\tau}{c}\sum_{l\leq k}\tau_{l}+\beta \sum_{s\leq r} h_s \tau_{l_s}}\rangle_{{{\btau}}} \\  \nonumber
D_W(\tau|h)&=& \langle
\rm e^{\frac{\beta}{2c}(\sum_{l\leq k}\tau_{l})^2+\frac{\beta\tau}{c}\sum_{l\leq k}\tau_{l}+\beta \sum_{s\leq r} h_s \tau_{l_s}}\rangle_{{{\btau}}},
\end{eqnarray}
with the short-hand $\langle f(\btau)\rangle_{\btau}=2^{-k}\sum_{\tau_1\ldots \tau_k=\pm 1} f(\tau_1,\ldots,\tau_k)$.
\newline
From  $P(M|k)$ we can deduce to what extent the network can perform extensive parallel retrieval, since the `pattern size' $k$ determines the associated overlap range via $-k\leq M \leq k$.

One observes that $W(h)=\delta(h)$ is a solution of (\ref{eq:single_eqn}) at any  noise level. If we inspect bifurcations of alternative solutions with nonzero moments $m_r=\int dh~ h^r W(h)$ (in particular with $m_1=0$ but $m_2\neq 0$, because $W(h)=W(-h)$), we find a second order transition along the critical surface in the $(\alpha,\beta,c)$ space defined by
\be\label{eqn:crit}
\alpha c^2
\sum_{k\geq 0}e^{-c}\frac{c^{k}}{k!}
\Big(
\frac{\int\! Dz~\tanh\Theta\cosh^{k+1}\Theta
}
{\int\! Dz~\cosh^{k+1}\Theta
}
\Big)^{\!2}=1,
\ee
where $\Theta=z\sqrt{\beta/c}\!+\!\beta/c$ and $Dz=(2\pi)^{-1/2}e^{-z^2/2}dz$. This expression is confirmed by the results of solving (\ref{eq:single_eqn})  via the population dynamics method  \cite{bethe}. The left-hand side of (\ref{eqn:crit}) obeys ${\rm LHS}\leq \alpha c^2$,  $\lim_{\beta\to 0}{\rm LHS}=0$ and $\lim_{\beta\to \infty}{\rm LHS}=\alpha c^2$.
Hence a transition at finite noise level $T_c=\beta_c^{-1}(\alpha,c) > 0$ to a new state with $W(h)\neq \delta(h)$ exists  as soon as $\alpha c^2>1$. The critical noise level goes to zero when $\alpha c^2=1$, i.e. at the percolation threshold. The transition line (\ref{eqn:crit})  is shown in the $(\alpha,T)$ plane in Fig. \ref{fig:Tc}.
\begin{figure*}[t]
\begin{center}
\includegraphics[width=.70\textwidth]{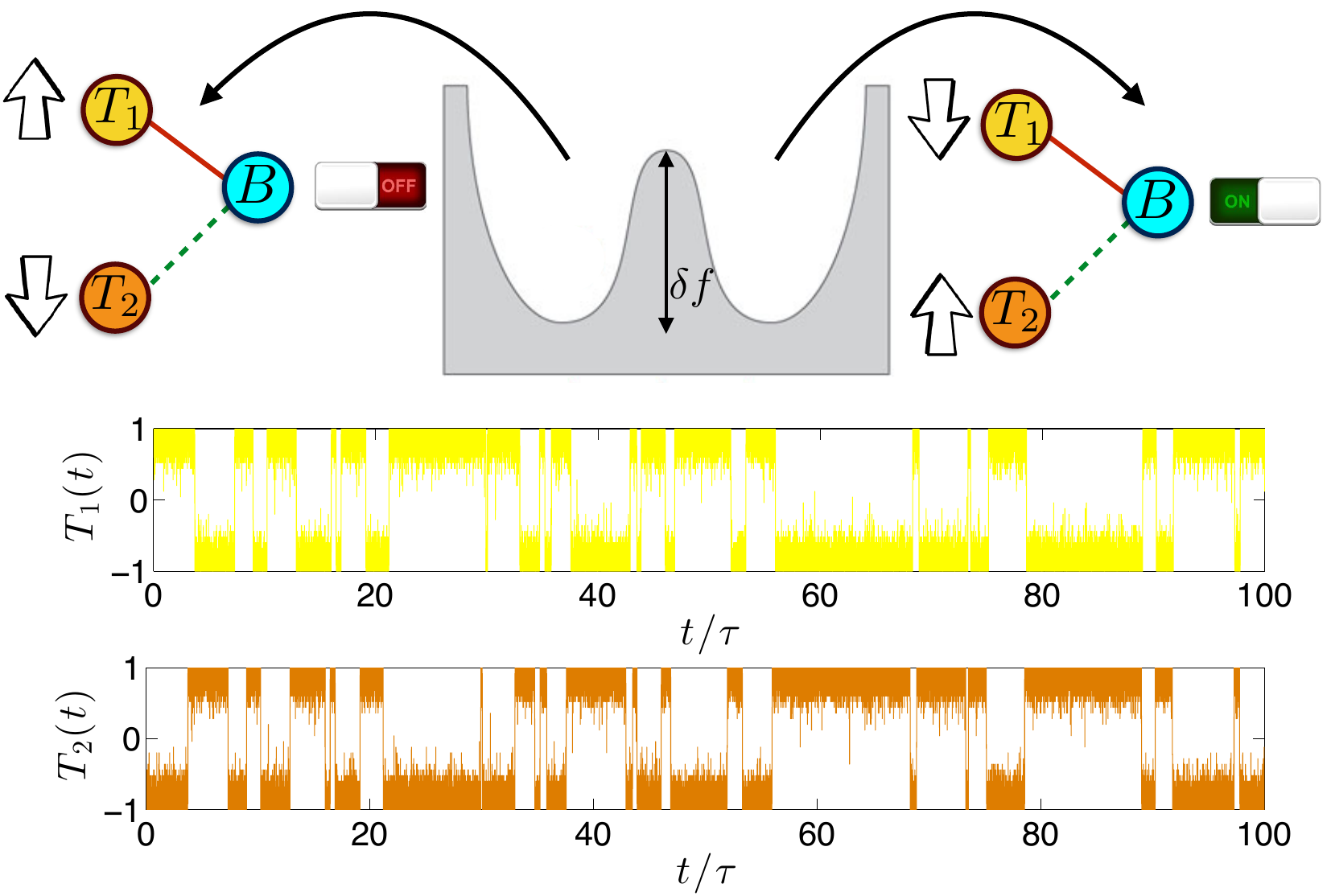}
\caption{Upper panel: Schematic representation of bi-stability induced by weak ergodicity breaking for a connected component of size $k=3$ in the bipartite interaction graph $\mathcal{B}$. In this example the component constitutes a {\em flip flop} \cite{SR1,SR2}, where two coordinator clones ($T_1$ and $T_2$) handle the expansion of an effector clone ($B$). Each clone is made of by $50$ cells. This system exhibits two free energy minima corresponding to $T_2$  and $B$ both active (while $T_1$ is quiescent) or to $T_2$ and $B_1$ both quiescent (while $T_1$ is active). The hopping rate between these states is $\tau \propto \exp(k \delta f)$, where $\delta f$ is the relative change in its intensive free energy. The time series for the magnetizations of the clones $T_1$ and $T_2$ are shown in the bottom:  note that, in the present context, time does not represent physical time, rather it solely counts the Monte Carlo steps. This component, upon marginalization over the effectors, is equivalent to a dimer in $\mathcal{G}$ where the two coordinators must be anti-parallel, and this constitutes a {\em logical clause} imposing that when one is firing the other is quiescent (and viceversa).} \label{fig:cinque}
\end{center}
\end{figure*}
In the under-percolated regime, i.e. for $\alpha c^2<1$, there is no possibility of a phase transition. Here the only solution of (\ref{eq:single_eqn}) is $W(h)=\delta(h)$, and (\ref{eqn:pm})
reduces to an expression corresponding to a Boltzmann distribution for a size-$k$ Curie-Weiss ferromagnet:
\be
P(M|k) = Z_k^{-1}\big\langle
\delta_{M,\sum_{l\leq k}\tau_l}~
\rme^{\beta(\sum_{l\leq k}\tau_l)^{2}/2c}
\big\rangle_{{{\btau}}}
\ee
 Hence for $\alpha c^2<1$ the cross-talk between different patterns vanishes. Each pattern effectively links to its own dedicated set of spins, and the system behaves as a set of $N_B$ disjunct  networks, each with a single stored pattern, and each acting as a {\em finite ferromagnet} (after the gauge transformation $\sigma_i\to\xi^{\mu}_i\sigma_i$). In the infinite noise limit $\beta\to 0$ we find the trivial $P(M|k)=\langle\delta_{M,\sum_{l\leq k}\tau_l} \rangle_{\btau}$, i.e. all  spins take random values. In the zero noise limit
 $\beta\to\infty$ we obtain $P(M|k)=\frac{1}{2}(\delta_{M,k}\!+\!\delta_{M,-k})$,
i.e. perfect retrieval.
Overall $P(M|k)$ goes from a single peak at $M=0$ for high noise levels, towards two symmetric peaks, at low noise levels; $P(M)$ always has a maximum at $M=0$.
Below the critical line in Fig. \ref{fig:Tc} the relevant solution of (\ref{eq:single_eqn}) has $W(h)\neq \delta(h)$. Now the effective Boltzmann factor in (\ref{eqn:pm})  acquires a further term  $\beta \sum_{s\leq r} h_s \tau_{l_s}$, which accounts for the fact  that the $N_B$ subsystems are no longer disconnected, leading to cross-talk interference via effective random fields $\{h_\ell\}$, which reduce the system's parallel processing ability. All our results are supported by numerical simulations \cite{long}. Note further that, as the percolation threshold is given by $\alpha c^2 =1$, assuming $c \sim O(1)$ (as experimentally suggested by the {\em chemical specificity} of cell's dialogues), the critical ratio for effectors vs coordinators $\alpha_c = [B]/[T] \sim 1$, again in plain agreement with the leukocytary formula (i.e. the immune system works properly when T cells are -of the same order but- more abundant than B cells \cite{janaway}).

\bigskip

Finally, it is important to stress that, since each subsystem (i.e. each clique as those sketched in Fig.$1$) is of {\em finite size} $k$, the system will exhibit only weak ergodicity breaking \cite{trap}, that is, free energy barriers between minima related to Hamiltonian (\ref{eq:reducedH}) do not diverge neither in the thermodynamic limit $N_T \to \infty$ (because, due to finite connectivity, they are proportional $\propto k$ and not to $\propto N_T$). This implies that the system may eventually jump spontaneously from one minimum to another -in the free energy landscape- corresponding to the two gauge symmetric magnetizations $M\!=\!\pm k$ (see Fig. \ref{fig:cinque}). Using $\delta f$ to label the (intensive) free energy (see again Fig. \ref{fig:cinque}), the typical time-scale for these stochastic transitions reads as $\tau \sim\rme^{k \delta f}$ (which tends to infinity only at the pathological zero noise level $T=\beta^{-1}$), and grows exponentially with the size $k$ of the subsystem (note that here time is meant solely in terms of Monte Carlo steps). These bi-stabilities are  due to intrinsic {\em small system's fluctuations} and are object of intense research at present \cite{cicli,hat,bistability,quorum,noi2016}. These may in fact have deep implications in homeostasis: beyond standard apoptotic pathways (e.g. via death Fas-like receptors \cite{apoptosi}), also a persistent lack of signalling  could prompt cellular depletion or functional reduction (i.e., cells that are not triggered within a given time-scale may undergo anergetic \cite{goodnow2} or apoptotic \cite{janaway} pathways) hence switching between positive and negative instructions to clones may shape opportunely their relative sizes.

In conclusion, we have shown how new insights and techniques from graph theory and statistical mechanics of finitely connected spin systems allow us to deepen our understanding of important aspects of the adaptive immune system, namely its remarkable and crucial ability to manage an extensive number of clones in parallel, and its possible relation to homeostatic regulation.


\vspace{1cm}

{\bf Acknowledgements}

The Authors are grateful to Gruppo Nazionale per la Fisica Matematica (GNFM-INDAM), trough Progetto-Giovani Agliari2016 and Progetto-Giovani Tantari2016, to Salento University and to the UK's  Biotechnology and Biological Sciences Research Council for financial support.


\begin{thebibliography}{9}
 

\bibitem{zerox} I.R. Cohen (Ed), {\em Theories of Immune Networks}, Springer-Verlag, New York (1988).

\bibitem{zerozero} M.A. Nowak, R.M. May, {\em Virus dynamics: mathematical principles of immunology and virology}, Oxford Univ. Press (2000).

\bibitem{zerouno} A.S. Perelson, G. Weisbuch, {\em Immunology for physicists}, Rev. Mod. Phys. \textbf{69}(4), 1219 (1997).

\bibitem{zerodue} R.J. De Boer, A.S. Perelson, {\em Size and connectivity as emergent properties of a developing immune network}, J. Theor. Biol. \textbf{149}(3), 381, (1991).

\bibitem{zerotre} R.J. De Boer, L.A. Segel, A.S. Perelson, {\em Pattern formation in one-and two-dimensional shape-space models of the immune system}, J. Theor. Biol. \textbf{155}(3), 295, (1992).

\bibitem{parisi} G. Parisi, {\em A simple model for the immune network}, Proc. Natl. Acad. Sci. USA \textbf{87}, 429, (1990).

\bibitem{JTB1} E. Agliari, et al., {\em A thermodynamic perspective of immune capabilities}, J. Theor. Biol.  \textbf{287}, 48, (2011).

\bibitem{JTB2} E. Agliari, et al., {\em Anergy  in self-directed B-cells from a statistical mechanics perspective}, J. Theor. Biol. \textbf{375}, 21, (2015).

\bibitem{prlnoi} E. Agliari, et al., {\em multitasking associative networks}, Phys. Rev. Lett. \textbf{109}, 268101, (2012).

\bibitem{deem} M.W. Deem, H.Y. Lee, {\em Sequence space localization in the immune system response to vaccination and disease}, Phys. Rev. Lett. \textbf{91}, 068101, (2003).

\bibitem{silvia1} S. Bartolucci, A. Annibale, {\em Associative networks with diluted patterns: dynamical analysis at low and medium load}, J. Phys. A: Math. Gen. \textbf{47}, 41, (2014).

\bibitem{silvia2} S. Bartolucci, A. Annibale, {\em A dynamical model of the adaptive immune system}, JSTAT P08017 (2015).

\bibitem{kosmir1} A. Ko\v{s}mrlj, et al., {\em  Thymic selection of T-cell receptors as an extreme value problem}, Phys. Rev. Lett.  \textbf{103}, 068103, (2009).

\bibitem{kosmir2} A. Ko\v{s}mrlj, et al., {\em  Quorum sensing allows T cells to discriminate between self and nonself}, Proc. Natl. Acad. Sci. USA \textbf{105}, 16671, (2008).

\bibitem{bialek} T. Mora, et al.,  {\em Maximum entropy models for antibody diversity}, Proc. Natl. Acad. Sci. USA \textbf{107}, 5405, (2010).

\bibitem{ton6} T. Uezu, C. Kadano, J.P.L.  Hatchett, A.C.C.  Coolen,  {\em  A large scale dynamical system immune network model with finite connectivity}, Prog. Theor. Phys. \textbf{161},
 385, (2006).

\bibitem{smallworld} E. Agliari, A. Barra,{\em A Hebbian approach to complex network generation}, Europhys. Lett. \textbf{94}, 10002, (2011).

\bibitem{barabasi} R. Albert, A.L. Barabasi, {\em Statistical mechanics of complex networks}, Rev. Mod. Phys. \textbf{74}, 47, (2002).

\bibitem{prlTon} J. P. L. Hatchett, I. Perez Castillo, A. C. C. Coolen,  N. S. Skantzos, {\em Dynamical replica analysis of disordered Ising spin systems on finitely connected random graphs}, Phys. Rev. Lett. \textbf{95}, 117204, (2005).

\bibitem{ton3} N.S. Skantzos, A.C.C.  Coolen, {\em $(1+\infty)$-dimensional attractor neural networks}, J.Phys. A: Math. Gen. \textbf{33}, 5785, (2000).

\bibitem{ton4} B. Wemmenhove, A.C.C. Coolen, {\em Finite connectivity attractor neural networks}, J. Phys. A: Math. Gen. \textbf{36}, 9617, (2003).

\bibitem{janaway} C. Janeway, P. Travers, M. Walport, M. Shlomchik, {\em Immunobiology}, Garland Science Publishing, New York, (2005).

\bibitem{Ton} A.C.C. Coolen, R. K\"{u}hn, P. Sollich, {\em Theory of Neural Information Processing Systems}, Oxford Press, Oxford, (2005).

\bibitem{hierarchical1} P. Moretti, M.A. Munoz, {\em Griffiths phases and the stretching of criticality in brain networks}, Nature Comm. \textbf{4}, 2521, (2013).

\bibitem{hierarchical2} E. Bullmore, O. Sporns, {\em Complex brain networks: graph theoretical analysis of structural and functional systems}, Nat. Rev. Neurosci. \textbf{10}(3), 186, (2009).

\bibitem{AACTold} E. Agliari, et al., {\em Immune networks: Multitasking capabilities at medium load}, J. Phys. A: Math. Gen. \textbf{46}, 335, (2013).

\bibitem{thompson} C.J. Thompson, {\em Mathematical Statistical Mechanics}, Princeton Univ. Press (1967).

\bibitem{noi2016} E. Agliari, et al., {\em Complete integrability of information processing by biochemical ractions}, Nature Sci. Rep. \textbf{6}, 36314 (2016).

\bibitem{SR1} E. Agliari, et al.,  {\em Notes on stochastic (bio)-logical gates: computing with allosteric cooperativity}, Nature Sci. Rep. \textbf{5}, 9415, (2015).

\bibitem{SR2} E. Agliari, et al.,  {\em Collective Behaviours: from biochemical kinetics to electronic circuits}, Nature Sci. Rep. \textbf{3}, 3458, (2013).

\bibitem{peter-prl} A. Annibale, et. al., {\em Extensive parallel processing on scale free networks}, Phys. Rev. Lett. \textbf{113}, 238106 (2014).

\bibitem{gerarchico-prl} E. Agliari, et. al., {\em Retrieval capabilities of hierarchical networks: From Dyson to Hopfield}, Phys. Rev. Lett. \textbf{114}, 028103, (2015).

\bibitem{long} E. Agliari, et al., {\em Immune networks: Multitasking capabilities close to saturation}, J. Phys. A: Math. Gen. \textbf{46}, 415003, (2013).

\bibitem{AGS} D. J. Amit, H. Gutfreund, H. Sompolinsky, {\em Storing infinite numbers of patterns in a spin-glass model of neural networks}, Phys. Rev. Lett. \textbf{55}, 1530, (1985).

\bibitem{jaynes} E.T. Jaynes, {\em Information theory and statistical mechanics}, Phys. Rev. \textbf{106}(4), 620 (1957).

\bibitem{bialek1} W. Bialek, {\em Biophysics: searching for principles}, Princeton Univ. Press (2012).

\bibitem{bialek2} E. Schneidman, et al., {\em Weak pairwise correlation imply strongly correlated network states in a neural population}, Nature \textbf{440}(7087), 1007, (2006).

\bibitem{trap} R.A. Denny, D.R. Reichman, J.P. Bouchaud, {\em Trap models and slow dynamics in supercooled liquids}, Phys. Rev. Lett. \textbf{90}, 025503, (2003).

\bibitem{goodnow2} C.C. Goodnow, {\em Cellular and genetic mechanisms of self tolerance and autoimmunity}, Nature \textbf{435}, 590, (2005).

\bibitem{bethe} M. Mezard, G. Parisi, {\em The Bethe lattice spin glass revisited}, Eur. Phys. J. B \textbf{20}, 217, (2001).

\bibitem{cicli} M. Samoilov, et al., {\em Stochastic amplification and signaling in enzymatic futile cycles through noise-induced bistability with oscillations}, Proc. Natl. Acad. Sci. USA \textbf{102}, 2310, (2005).

\bibitem{hat} T.  Lipniacki, et al., {\em Stochastic effects and bistability in T cell receptor signaling}, J. Theor. Biol. \textbf{254}, 110, (2008).

\bibitem{bistability} M.N. Artyomov, et al., {\em Purely stochastic binary decisions in cell signaling models without underlying deterministic bistabilities}, Proc. Natl. Acad. Sci. USA \textbf{104}, 18958, (2007).

\bibitem{quorum} T.C. Butler, et al., {\em Quorum sensing allows T cells to discriminate between self and nonself} Proc. Natl. Acad. Sci. USA \textbf{110}, 11833, (2013).

\bibitem{apoptosi} G. Wu, Y. Shi, {\em Apoptosis signaling pathways and lymphocyte homeostasis}, Nature Cell Research \textbf{17}, 759, (2007).
%
%


\end{thebibliography}
\end{document}